\documentclass{article}

\usepackage{PRIMEarxiv}
\usepackage{amsmath}
\usepackage[utf8]{inputenc} % allow utf-8 input
\usepackage[T1]{fontenc}    % use 8-bit T1 fonts
\usepackage{hyperref}       % hyperlinks
\usepackage{url}            % simple URL typesetting
\usepackage{booktabs}       % professional-quality tables
\usepackage{amsfonts}       % blackboard math symbols
\usepackage{nicefrac}       % compact symbols for 1/2, etc.
\usepackage{microtype}      % microtypography
\usepackage{lipsum}
\usepackage{fancyhdr}       % header
\usepackage{graphicx}       % graphics
\graphicspath{{media/}}     % organize your images and other figures under media/ folder

\title{Hierarchical Causal Uplift Modeling in Overlapping Customer Journeys
}

\author{
  Jorge Pellegrini \\
  Marketing Science, Despegar \\
  Buenos Aires, Argentina \\  
  \texttt{jpellegrini@fra.utn.edu.ar} \\
  ORCID: \href{https://orcid.org/0000-0003-0526-7743}{0000-0003-0526-7743}
}

\begin{document}
\maketitle

\begin{abstract}
Digital travel platforms often operate multiple marketing journeys simultaneously, 
resulting in overlapping user exposures that bias the standard A/B lift estimation. 
Because traditional lift experiments assume treatment isolation, the observed lifts 
reflect only marginal effects and may substantially underestimate the total incremental 
impact of each journey.

This work introduces a Hierarchical Causal Lift Model that decomposes pure and global effects under journey overlap. Each journey is modeled as a multiplicative causal factor, and the interaction terms capture potential synergies or cannibalizations. The model is estimated through a Monte Carlo framework that incorporates uncertainty in overlap proportions, observed lifts, and single-journey effects. Regularized non-linear least squares are complemented with Monte Carlo simulation to quantify parameter uncertainty and assess the robustness of the solution.

Applied to an active user base of approximately three million users, the model reveals 
positive but modest synergies between journeys and shows that pure lifts are significantly larger than those observed experimentally. The predicted global lift closely matches the experimentally measured value, demonstrating the ability of the model to recover incremental effects in an interpretable manner.
\end{abstract}

% keywords can be removed

\section{Introduction}
Travel e-commerce platforms operate in an increasingly competitive and fast-evolving market, where user attention is scarce, and users can easily switch between competing platforms\cite{bergemann2024data}. A central challenge in this ecosystem is maintaining high levels of customer engagement and conversion across multiple channels and devices\cite{asante2024leveraging}. To address this, companies deploy automated marketing journeys, defined as communication flows that are triggered when a user meets specific behavioral conditions (for example, reaching a shopping cart without completing a purchase)\cite{dinu2020using,rolando2025marketing}. Each journey automatically delivers a sequence of personalized messages (such as email, push, in-app, or re-targeting) intended to reengage the user and drive conversion, often adapting in timing and content based on observed user responses.

The effectiveness of these journeys is typically assessed through A/B lift experiments\cite{siroker2015b}. At the start of an experiment, users are randomly assigned to one of two branches: a treatment group that receives the journey as usual and a control group that does not receive any communication. After a sufficient observation period, the difference in conversion rates between both groups represents the lift attributable to that specific journey. This design facilitates transparent governance and enables a clear interpretation of the results. Thus, A/B testing remains the standard approach for evaluating the isolated performance of individual marketing flows\cite{kohavi2009controlled,kohavi2020trustworthy}.

However, real-world experimentation presents structural limitations. Users can be simultaneously exposed to multiple journeys during the same measurement window, leading to interaction effects that complicate the interpretation of observed lifts\cite{anderl2016mapping}. Journeys may overlap in eligibility rules, timing, and audience segments, producing nonlinearities that standard holdout designs do not capture\cite{li2014attributing}. Moreover, running simultaneous A/B tests across all journeys would require suppressing approximately half of total communications, a scenario that is neither operationally feasible nor business-justifiable due to revenue risk and customer experience considerations. Therefore, the assumption of independence between journeys rarely holds\cite{lewis2014online}. Consequently, relying only on the marginal lift of each journey may underestimate or misattribute the total incremental impact of the overall experimentation system.

To address these challenges, we propose a hierarchical causal model that recovers both individual and global lift effects under overlapping exposures. The model represents each journey as a multiplicative causal factor and explicitly accounts for interactions arising from concurrent activations. Using the observed lifts and exposure patterns, the approach provides corrected estimates that reflect the incremental impact of each journey within the broader experimentation system.

\section{Materials and Methods}
\label{sec:Materials and Methods}

This section describes the journeys included in the analysis, the considerations applied when computing lifts, and the proposed methodological framework.

\subsection{Active User Base}

The analysis focuses on the active user base, defined as users who performed specific actions on Despegar’s platform within a given time window and satisfied additional filters related to location and site activity. All experiments were restricted to this population.

\subsection{Journeys Description}

Three journeys for this active user base were analyzed:

\begin{itemize}
\item Search Journey: triggered when a user performs browsing or search actions on the site, without necessarily progressing to the purchase flow.
\item Flow Journey: activated when a user reaches a specific stage in the purchase flow (for example, adding items to the cart) but does not complete the transaction. 
\item Offer Journey: triggered when a user becomes eligible to receive specific promotional communications.
\end{itemize}

These are referred to as the Search, Flow, and Offer journeys, respectively.

\subsection{Lift Experiments}

Each journey was tested through a dedicated A/B experiment at different points in time, such that the corresponding experimental periods did not overlap.

A fixed random seed was used to define the control and treatment branches.
This ensures, for example, that a user assigned to the control branch for the Search journey (who does not receive communications) will also belong to the same control branch in the Flow or Offer experiments.
This design maintains consistency across experiments and allows for measuring lifts over comparable user populations.

Importantly, the lift metric does not depend on whether the journey occurred before or after a purchase event.
The experiment measures differences in conversion across the entire active user base, rather than only among users who entered each journey.
This design provides a more business-oriented measure of incremental impact, since it avoids conditioning on the specific activation criteria of each journey and supports the modeling strategy developed in subsequent sections.

Using the fixed random seed, the active base is split into two disjoint branches:

\begin{itemize}
\item Branch A (control): users who, if exposed to a journey, would fall into the group that does not receive communication.
\item Branch B (treatment): users who, if exposed, would fall into the group that receives communication. 
\end{itemize}

In addition to individual journey tests, a global lift analysis was performed.
This analysis applies the A/B framework to the simultaneous activation of all journeys, providing an estimate of the overall incremental effect of the full system.
The resulting global lift provides an empirical benchmark for evaluating system-level effects. The computation of the lift for each experiment and the quantification of the associated uncertainty are described in the following section.

\subsection{Regression Framework}
\label{sec:regression}
The A/B experiment for the journey $j$ is formalized by defining two branches of the active user base\cite{kohavi2013online}:

\begin{itemize}
    \item a \textbf{test group} with $n_T$ users, of whom $m_T$ completed a purchase.
    \item a \textbf{control group} with $n_C$ users, of whom $m_C$ completed a purchase.
\end{itemize}

The empirical conversion rates for each branch are therefore:
\begin{equation}
\hat{p}_T = \frac{m_T}{n_T}, \qquad
\hat{p}_C = \frac{m_C}{n_C}
\end{equation}

To express this structure within a regression framework, we define for each user $i$:

\[
\begin{array}{rclcrcl}
Y_i &=& 
\left\{
\begin{array}{ll}
1, & \text{if the user converts},\\
0, & \text{otherwise},
\end{array}
\right.
&\qquad&
T_i &=&
\left\{
\begin{array}{ll}
1, & \text{if the user belongs to the control group},\\
0, & \text{if the user belongs to the test group}.
\end{array}
\right.
\end{array}
\]

The expected conversion rate is then modeled as:
\begin{equation}
\mathbb{E}[Y_i] = \beta_0 + \beta_1 T_i
\end{equation}

where the coefficients correspond to the population-level conversion rates:
\begin{equation}
\beta_0 = \mathbb{E}[Y_i \mid T_i = 0] = p_T, 
\qquad 
\beta_0 + \beta_1 = \mathbb{E}[Y_i \mid T_i = 1] = p_C
\end{equation}

We estimate the coefficients $\hat{\beta}_0$ and $\hat{\beta}_1$ using 
ordinary least squares (OLS) for all the users in the experiment. 
This approach provides unbiased estimates of the mean conversion rate in each branch\cite{angrist2009mostly}: 
the intercept $\hat{\beta}_0$ corresponds to the conversion rate of the test group, 
while the sum $\hat{\beta}_0 + \hat{\beta}_1$ represents that of the control group.

To quantify the uncertainty associated with the estimated lift, we apply the \textit{Delta Method}\cite{CasellaBerger2024}, a first-order Taylor approximation that allows deriving the variance of a non-linear transformation of estimated parameters. 

The estimated lift is defined as the relative difference between both branches\cite{deng2013improving}:

\begin{equation}
L = g(\hat{\boldsymbol{\beta}}) = \frac{\hat{\beta}_1}{\hat{\beta}_0}.
\label{eq:lift_function}
\end{equation}

If $\hat{\boldsymbol{\beta}} = (\hat{\beta}_0, \hat{\beta}_1)$ follows an approximately 
bivariate normal distribution with the covariance matrix $\Sigma_{\hat{\boldsymbol{\beta}}}$, 
then the variance of $L$ can be approximated as:

\begin{equation}
\mathrm{Var}(L) \approx 
\nabla g(\hat{\boldsymbol{\beta}})^{\top} 
\, \Sigma_{\hat{\boldsymbol{\beta}}} \,
\nabla g(\hat{\boldsymbol{\beta}}),
\label{eq:delta_var}
\end{equation}

where the gradient of the transformation $g(\hat{\boldsymbol{\beta}})$ with respect to 
$\hat{\boldsymbol{\beta}}$ is given by:

\begin{equation}
\nabla g(\hat{\boldsymbol{\beta}}) = 
\left[
\frac{\partial L}{\partial \beta_0}, 
\frac{\partial L}{\partial \beta_1}
\right]_{\beta=\hat{\beta}}
=
\left[
-\frac{\hat{\beta}_1}{\hat{\beta}_0^2}, \;
\frac{1}{\hat{\beta}_0}
\right].
\label{eq:delta_grad}
\end{equation}

Substituting the estimated covariance matrix of the coefficients, obtained from the regression output, 
into Equation~\ref{eq:delta_var} yields the estimated variance of the lift $\widehat{\mathrm{Var}}(L)$.  
The corresponding standard error is then $\widehat{\mathrm{SE}}(L) = \sqrt{\widehat{\mathrm{Var}}(L)}$, 
and the 95\% confidence interval can be computed as:

\begin{equation}
CI_{95\%} = 
\left[
L - 1.96 \times \widehat{\mathrm{SE}}(L), \;
L + 1.96 \times \widehat{\mathrm{SE}}(L)
\right].
\end{equation}

This analytical approximation provides a computationally efficient and interpretable measure 
of the uncertainty of the lift.

\subsection{Limitations of the Observed Lift and Motivation for the Model}

Although the observed lift estimated through A/B testing provides an unbiased measure of the incremental effect
of a single journey under isolated conditions, it does not capture the full potential impact of that journey
within the broader experimentation ecosystem.

In practice, users can interact simultaneously or sequentially with multiple active journeys. 
As a result, the observed lift of a given journey reflects only the marginal effect 
measured within the specific experimental configuration, 
and not its total contribution to overall user conversion. 
Interactions between journeys can generate both \textit{synergies} (reinforcing effects) 
and \textit{cannibalizations} (dampening effects), 
depending on timing, targeting, and user heterogeneity.

To address these limitations, we propose a probabilistic model 
that explicitly accounts for the overlap structure among journeys. 
The key idea is to simulate, via Monte Carlo sampling, 
the probability that users belong to one or more journey combinations 
and to use these simulated joint distributions as inputs 
for reconstructing the observed lifts.

Under this framework, the observed lifts are treated as constraints 
that emerge from combinations of \textit{pure lifts} 
(the effect of each journey in isolation) 
and interaction factors representing synergies or cannibalizations. 
By solving an optimization problem that minimizes the deviation between
the simulated and empirical lifts, 
the model allows recovery of both the \textit{pure} and the \textit{global} lifts 
consistent with the observed experimental data.

The following section presents the proposed model formulation, 
its mathematical structure, and the simulation-based estimation procedure.

\subsection{Model Formulation}

The proposed model aims to recover the pure and total lift effects of multiple overlapping journeys. 
The key assumption is that each journey contributes multiplicatively to the overall probability of conversion 
and that the joint effect of several active journeys can be represented through interaction factors 
capturing potential synergies or cannibalizations. 
The journeys under analysis are \textit{Search (S)}, \textit{Flow (F)}, and \textit{Offer (O)}.

\subsubsection{Model structure}

Each journey $j \in \{S, F, O\}$ is associated with a multiplicative factor 
$f_j = 1 + \ell_j$, where $\ell_j$ represents the net lift of that journey when active. 

Beyond individual effects, pairwise and higher-order interactions are introduced 
through the parameters $\kappa_{SF}$, $\kappa_{SO}$, $\kappa_{FO}$, and $\kappa_{SFO}$. 
A value of $\kappa = 1$ indicates that there is no interaction (pure independence), 
while deviations from 1 represent amplification or attenuation between journeys.

For any combination of journeys $c$, the total effect is expressed as the product 
of individual and interaction factors:

\begin{equation}
F(c) = 
\prod_{j \in c} f_j \times 
\kappa_{SF}^{[\{S,F\} \subseteq c]} \,
\kappa_{SO}^{[\{S,O\} \subseteq c]} \,
\kappa_{FO}^{[\{F,O\} \subseteq c]} \,
\kappa_{SFO}^{[\{S,F,O\} \subseteq c]}.
\end{equation}

\subsubsection{Population-level expected effects}

Let $p_c$ denote the empirical probability that users belong to the combination $c$ 
(estimated from the proportions of observed overlap). 

The average population effect when all journeys are active is then:
\begin{equation}
E_{\text{on}} = \sum_c p_c F(c).
\end{equation}

If a given journey $j$ is turned off, its corresponding factor and interaction terms 
are removed, resulting in:
\begin{equation}
E_{\text{off}(j)} = 
\sum_c p_c \frac{F(c)}{f_j^{[j \in c]} 
\prod_S \kappa_S^{[\;S \subseteq c, \; j \in S\;]}}.
\end{equation}

The lift observed when the journey is deactivated $j$ is defined as:
\begin{equation}
L^{\text{obs}}_{\text{on}(j)} = 
\frac{E_{\text{on}}}{E_{\text{off}(j)}} - 1.
\end{equation}

\subsubsection{Estimation problem and optimization}

The goal is to estimate the parameter vector:
\begin{equation}
\theta = (f_S, f_F, f_O, \kappa_{SF}, \kappa_{SO}, \kappa_{FO}, \kappa_{SFO})
\end{equation}
such that the predicted lifts of the model reproduce the experimental lifts 
observed $L^{\text{obs}}_{\text{on}(j)}$.

The residuals for each journey are defined as follows:
\begin{equation}
r_j = 
\left( 
\frac{E_{\text{on}}}{E_{\text{off}(j)}} - 1
\right)
- L^{\text{obs}}_{\text{on}(j)}.
\end{equation}

To ensure stable estimation and prevent overfitting of interaction parameters, 
a regularization term is added that penalizes deviations of interaction factors 
from neutrality ($\kappa = 1$)\cite{hoerl1970ridge}:
\begin{equation}
r_{\kappa}^{\text{ridge}} = 
\sqrt{\lambda_{\text{inter}}} \, (\kappa - 1),
\end{equation}
where $\lambda_{\text{inter}}$ controls the strength of the penalty. 

In addition to the ON/OFF experimental lifts, the model incorporates the so-called 
pure lifts, defined as the incremental effects measured on users exposed exclusively 
to a single journey. Because these estimates are noisy and reflect smaller and more 
selective subpopulations, they are not imposed as hard constraints. Instead, they are 
included as soft anchors by adding a set of residual terms of the form 
$(f_j - 1) - \ell^{\text{pure}}_j$, weighted by a tuning parameter $w_{\text{pure}}$. 
This formulation corresponds to a Gaussian prior on the main effects, stabilizing 
the estimation under overlap and preventing degenerate solutions while allowing deviations 
when strongly supported by the data \cite{gelman1995bayesian}.

The full objective function combines residuals and regularization:
\begin{equation}
\min_{\theta} \sum_i r_i(\theta)^2.
\end{equation}

Minimizing this objective is equivalent to performing a 
Gaussian maximum a posteriori estimation, 
under the assumption that the residuals are approximately normal \cite{m2006pattern}. 
The optimization is solved numerically via a non-linear least squares procedure, 
and the resulting parameters allow reconstructing both 
\textit{pure lifts} and the \textit{global combined lift} 
consistent with the observed experimental data.

Two estimation components were employed in this work: 
a deterministic non-linear least-squares optimization and a Monte Carlo simulation procedure. 
The least-squares method estimates the parameter vector by minimizing the squared residuals 
between the observed and model-predicted lifts, while the Monte Carlo component quantifies 
the uncertainty induced by variability in journey overlap proportions and experimental lift measurements.

\subsubsection{Monte Carlo Simulation}
This procedure consists of modeling uncertainty in the population proportions associated with each journey combination segment by sampling from a Dirichlet distribution.
In parallel, the observed ON/OFF lifts and the pure lifts are perturbed by drawing from their empirical uncertainty distributions. 
Given these sampled quantities, the non-linear least-squares solver is applied to obtain 
a point estimate of the model parameters for that simulation draw.

Repeating this process over many iterations yields an empirical distribution for the parameters, 
reflecting the combined uncertainty arising from the overlap of the journeys, experimental noise, 
and the soft constraints imposed by pure lifts\cite{robert1999monte}. 
Posterior summaries (mean, median and percentile ranges) provide stable estimates of 
main effects ($f_S, f_F, f_O$), interaction factors ($\kappa_{SF}, \kappa_{SO}, \kappa_{FO}, \kappa_{SFO}$), 
and derived quantities such as pure and global lifts\cite{gelman1995bayesian}.

This simulation-based approach not only captures parameter uncertainty, but also validates the robustness of the deterministic solution, since convergence patterns across Monte Carlo iterations indicate that the optimization landscape behaves well and that estimates are identifiable under realistic data variability\cite{hastie2009elements}.

\section{Results}

The active user base consists of approximately three million users, 
evaluated over regular and comparable time windows for each lift experiment. Monte Carlo estimation was performed using 5{,}000 iterations, with interaction 
regularization set to $\lambda_{\text{inter}} = 0.30$ and a soft-anchor weight of 
$w_{\text{pure}} = 0.20$ for the pure-lift constraints. 
The model estimated the following mean interaction effects (synergies) between journeys, 
expressed as percentage changes relative to the independent expectation:

\[
\begin{aligned}
\text{Synergy}_{S,F} &: \text{mean } = 0.13\% \\
\text{Synergy}_{S,O} &: \text{mean } = 0.02\% \\
\text{Synergy}_{F,O} &: \text{mean } = 0.11\% \\
\text{Synergy}_{S,F,O} &: \text{mean } = 0.09\%
\end{aligned}
\]

All interaction terms are positive on average, indicating small but consistent synergies across journeys. 
The relatively low value of the three-way interaction (\textit{S--F--O}) is primarily associated with the low probability of this simultaneous exposure occurring in the population. Moreover, when users are exposed to all three journeys, the incremental gain in conversion is not particularly 
pronounced, suggesting that such users are already highly engaged and that additional communications provide 
limited marginal benefit. In contrast, the stronger \textit{F--O} relationship indicates that the offer system generates a 
significant impact when the user has reached an advanced stage of the purchase flow. 
This pattern suggests that users who are already close to conversion are more likely to respond positively 
to offer-based communications, even beyond the influence of search-related activations (\textit{S--O}). Finally, the lower synergy observed between \textit{S--F},although it might intuitively be expected to be high,
implies that users at this stage are already in a decision-making process, and the simultaneous activation of 
both journeys does not produce a clearly incremental effect. 
This result can be interpreted by noting that the messages from the Search and Flow journeys, 
although distinct in style, tend to trigger a similar behavioral response, thereby limiting 
their combined incremental impact.

For the Offer journey, the estimated pure lift is approximately four times the magnitude of the observed experimental lift. For the Search journey, the estimated effect is about 
4.6 times the observed value, while for the Flow journey the modeled lift is roughly 2.5 times larger than its experimental counterpart. These proportional differences reflect the attenuation present in the observed lifts due to overlap among journeys. Once such interference is removed by the model, the underlying incremental 
effects become substantially clearer, particularly for journeys that are 
more frequently co-activated with others.

With respect to the total lift predicted by the model, the estimated value represents 
approximately 85\% of the experimentally measured global lift. 
This difference is largely explained by the fact that the model does not incorporate 
post–purchase service interactions, which contribute to the observed global effect. 
Even with this omission, the model achieves a favorable level of performance and 
successfully captures the majority of the incremental impact observed in practice.

\section{Conclusion}
This work presents a hierarchical causal framework for estimating the incremental 
effects of multiple overlapping marketing journeys in a large-scale digital travel 
platform. Traditional A/B lift experiments, while effective for evaluating isolated 
journeys, do not account for concurrent exposures that arise naturally in real user 
behavior. As a result, observed experimental lifts capture only marginal effects and 
may substantially underestimate the true incremental impact of each journey.

The proposed Monte Carlo–based model addresses this limitation by jointly estimating 
individual journey effects and interaction terms under uncertainty in exposure 
patterns, experimental noise, and pure-lift variability. The method produces stable 
parameter estimates and provides an interpretable decomposition of global performance 
into pure and interaction components. The estimated pure lifts are substantially larger 
than the observed experimental lifts, reflecting the attenuation introduced by journey 
overlap, while the interaction terms indicate small but consistent synergies between 
journeys.

The global lift predicted by the model closely matches the total lift measured 
experimentally, demonstrating that the framework captures the majority of incremental 
impact at the system level. Overall, this approach offers a principled, data-efficient, 
and privacy-preserving methodology for evaluating the performance of multi-journey 
orchestration. It provides actionable insights for the design, prioritization, and 
optimization of marketing journeys in complex digital environments.

%Bibliography
\bibliographystyle{unsrt}  
\bibliography{references}

\end{document}